\documentclass[11pt,english]{achemso}
\usepackage[T1]{fontenc}
\usepackage[latin9]{inputenc}
\usepackage{geometry}
\usepackage{amstext}
\usepackage{graphicx}
\usepackage{setspace}
\usepackage{wasysym}
\makeatletter
\setkeys{acs}{articletitle = true}

\usepackage{amstext}

\title{On the elasticity of polymer model networks containing finite loops}
\author{Michael Lang}
\affiliation{Leibniz Institute of Polymer Research Dresden, Hohe Stra?e 6, 01069
Dresden, Germany}
\email{lang@ipfdd.de}

\usepackage{babel}

\makeatother

\usepackage{babel}
\begin{document}
\begin{center}
\includegraphics[width=0.48\paperwidth]{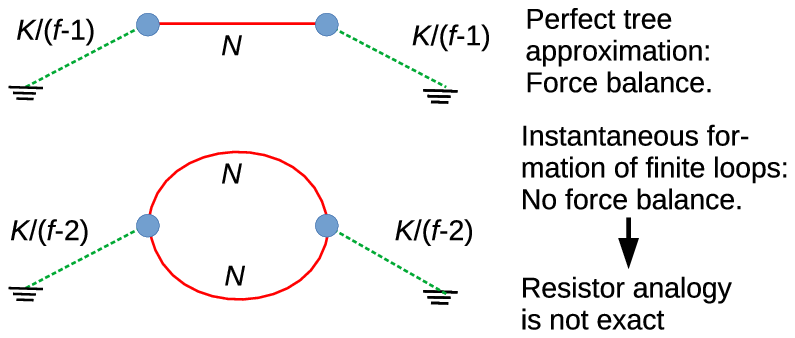}
\par\end{center}
\begin{abstract}
Based upon the resistor analogy and using the ideal loop gas approximation
(ILGA) it is shown that only pending loops reduce the modulus of an
otherwise perfect network made of monodisperse strands and junctions
of identical functionality. Thus, the cycle rank of the network with
pending structures removed (cyclic and branched) is sufficient to
characterize modulus, if the resistor analogy can be employed. It
is further shown that it is impossible to incorporate finite cycles
into a polymer network such that individual network strands are at
equilibrium conformations while maintaining simultaneously a force
balance at the junctions. Therefore, the resistor analogy provides
only an approximation for the phantom modulus of networks containing
finite loops. Improved approaches to phantom modulus can be constructed
from considering a force balance at the junctions, which requires
knowledge of the distribution of cross-link fluctuations in imperfect
networks. Assuming loops with equilibrium conformations and a force
balance at all loop junctions, a lower bound estimate for the phantom
modulus, $G_{\text{ph}}\approx\left(\xi-c_{\text{f}}L_{1}\right)kT/V$
is obtained within the ILGA for end-linked model networks and in the
limit of $L_{1}\ll\xi$. Here, $L_{1}$ is the number of primary (``pending'')
loops, $\xi$ the cycle rank of the network, $k$ the Boltzmann constant,
$V$ the volume of the sample, and $T$ the absolute temperature.
$c_{\text{f}}$ is a functionality dependent coefficient that is $\approx2.56$
for junction functionality $f=3$ and $\approx3.06$ for $f=4$, while
it converges quickly towards $\approx4.2$ in the limit of large $f$.
Further corrections to phantom modulus beyond finite loops are addressed
briefly. 
\end{abstract}

\section{Introduction}

Recent work has revived the discussion about the elasticity of rubber
networks or gels, as the detection of finite loops (also called ``finite
cycles'') inside networks became possible from the experimental side
\cite{Lange2011,Zhou2012}. This advance stimulated the development
of enhanced models for the elasticity of phantom networks that go
beyond the ideal tree approximation \cite{Zhong2016,Lang2018,Lin2019,Gusev2019}.
These models arrive at results that are distinct from classical work
that is typically discussed in textbooks of polymer physics, which
challenges our understanding of rubber elasticity.

One cornerstone of the original treatment of rubber elasticity is
that phantom modulus is equivalent to the cycle rank of the network
\cite{Flory1982}, if cycles in pending structures are neglected.
This has been argued previously using graph theory \cite{Flory1982},
using micro-networks \cite{Graessley1975}, or within the perfect
tree approximation of a perfect network where all strands are of equal
degree of polymerization and all junctions of same functionality \cite{Flory1976}.
Also, a numerical proof exists \cite{Higgs1988} based upon the resistor
analogy for polydisperse networks also within the perfect tree approximation
that indicates some generality of this result.

However, real networks are made entirely of finite loops and it has
been shown in recent work \cite{Zhong2016,Lang2018,Panyukov2019}
that these finite loops lead to a reduced contribution of the elastic
strands to the phantom modulus. These results were obtained in the
so-called ``ideal loop gas approximation'' (ILGA) where a single
loop resides inside an otherwise perfect tree as a first order approximation
for the impact of a finite loop on elasticity. These newer results
have been challenged by a very recent paper \cite{Lin2019}, that
proposes an intermediate result in between the classical work on phantom
modulus and the more recent work in Refs. \cite{Zhong2016,Lang2018}.
It was claimed that original works \cite{Flory1982,Higgs1988} are
incorrect regarding finite loops that consist of $i=2$ strands. Furthermore,
an additional force balance term was added to correct the resistor
analogy based estimate of elasticity, that was absent in earlier work
\cite{Zhong2016} or treated differently in other work \cite{Lang2018}.

These conflicting results are taken in the present paper as motivation
to discuss in more detail the principles for computing modulus in
phantom model networks made of finite cycles. This discussion addresses
several key points beginning with the resistor analogy, its limitations,
force balance conditions, and the implications when generalizing the
ILGA to a non-ideal network environment. It will be shown that original
works \cite{Flory1982,Higgs1988} are correct when applying the resistor
analogy to the ILGA. Unfortunately, this analytical result is only
approximate as the resistor analogy breaks down once finite loops
appear within the network since a force balance cannot be obtained
within a finite loop, if the loop and the surrounding network are
all made of strands at reference size. Furthermore, a lower bound
for the modulus is derived for end-linked model networks under the
consideration of the degree of polymerization (DP) distribution of
finite loops in a network. Finally, other corrections beyond finite
loops that are not considered in the classical work on phantom modulus
are addressed briefly.

\section{\label{sec:The-resistor-analogy}The resistor analogy}

\begin{figure}
\begin{centering}
\includegraphics[width=0.36\columnwidth]{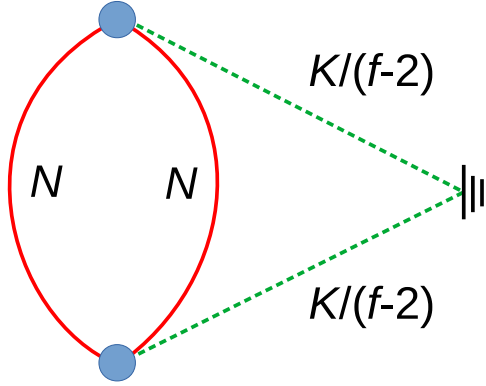}
\par\end{centering}
\caption{\label{fig:Resistor-analogue}Resistor analog for a loop made of $i=2$
strands of $N$ segments. The blue circles represent the junctions
of the loop. The virtual strands of $K/(f-2)$ describe the effective
connection to ground through the $f-2$ strands of $K$ effective
segments. See Ref. \cite{Lang2018} for more details.}
\end{figure}

Let us start with the resistor analogy for the ILGA. Below, we use
approach and notation of Ref. \cite{Lang2018} to double check the
results of Ref. \cite{Lin2019}. The basic structure that is discussed
is a single loop made of $i$ strands that is incorporated into an
otherwise perfect tree network. $N$ is the degree of polymerization
(DP) of a network strand, $f$ the junction functionality, and $K=(f-1)N/(f-2)$
the DP that describes the virtual strand running from a junction through
a single adjacent bond to the non-fluctuating elastic background.
Within the resistor analog, the DP's are treated as resistors $\propto N$.
Thus, $K/(f-2)$ is the DP (resistor) of the equivalent chain that
models the net connection through the $f-2$ remaining connections
to ground that run not trough the finite loop \cite{Lang2018}, see
Figure \ref{fig:Resistor-analogue} where the simplest example of
a loop of order $i=2$ is shown.

In the resistor analogue, one asks the question: What is the portion
of the current injected at two adjacent junctions (i.e. junctions
that are connected directly through a single chain) that runs through
a particular wire \cite{Higgs1988}? To answer this question, one
removes first the ``wire'' under consideration from the network
and computes then the conductance between the two junctions through
the remaining network structure. Here, this wire is one $N$-mer of
the loop $i=2$. The resulting conductance is modeled within the resistor
analogy by an effective strand of $N_{\text{eff}}$ segments, where
$N_{\text{eff}}$ is computed by treating it as a resistor. Thus,
the total conductance between the two junctions in Figure \ref{fig:Resistor-analogue}
is the sum of the conductances through the remaining $N$-mer and
through the two strands to ground: 
\begin{equation}
\frac{1}{N_{\text{eff}}}=\frac{1}{N}+\frac{f-2}{2K}.\label{eq:Ne}
\end{equation}
The contribution of the removed $N$-mer of the loop to the total
elastic strand between both junctions (after inserting it back into
the loop) is then given by the ratio between the $N$-mer and the
combined strand $N+N_{\text{eff}}$ as in the classical derivation
of the phantom modulus \cite{Rubinstein2003}. One obtains here 
\begin{equation}
\frac{N}{N+N_{\text{eff}}}=\frac{f-2}{f}+\frac{2}{f^{2}}.\label{eq:e2}
\end{equation}
This contribution is \emph{larger} than the ideal contribution by
the last term of $2/f^{2}$. The corresponding change in the elastic
contributions of the surrounding network strands were computed in
equation (S31) of the supporting information of Ref. \cite{Lang2018}
providing a net \emph{reduction} of the elastic contribution by $2/f^{2}$,
which \emph{cancels exactly }the enhancement inside the loop. Therefore,
there is no change in the \emph{sum} of all point-to-point conductances
between two adjacent junctions inside the resistor network when rewiring
the strands of a perfect network to form a single loop of order $i=2$.
Accordingly, the \emph{sum} of all effective elastic contributions
between any pair of directly connected junctions is not modified for
$i=2$.

The above result is readily generalized to $i\ge3$ with the approach
presented in the supporting information of Ref. \cite{Lang2018}.
Here, loops made of $i\ge3$ strands are subsequently reduced to an
asymmetric loop of two strands, where the corresponding strands to
ground, $K_{\text{av}}$ from equation (S21) of Ref. \cite{Lang2018},
replace the $K/(f-2)$ term above and the above ``bypass'' of $N$
segments, (i.e. the second strand that connects the pair of junctions
not through ground) is computed as $N_{\text{k}}$ in equation (S16)
and replaces $N$ in equation (\ref{eq:Ne}) above. These modifications
lead consistently to an exact cancellation of the enlarged elastic
contribution of the loop strands by the reduced contribution of the
surrounding network strands. The supporting information contains a
short FORTRAN code that performs the corresponding recursion for a
range of different $i$ and $f$. The last column of the output is
the net change in the point-to-point conductances of single wires
between all pairs of adjacent junctions when averaged over the whole
tree network with the loop in the middle. The first two columns of
the output are $f$ and $i$ respectively.

The above results are identical to the discussion in Ref. \cite{Lin2019}
for $i\ge3$ and differ only for $i=2$ in the contribution of the
two strands that form the loop. The difference is that above, these
strands were treated independently, while these were first combined
into a single strand of half the DP in Ref. \cite{Lin2019} but never
split again when considering the contributions of all real network
strands. Since the combination into one strand removes one elastic
strand from the network, it was consistently found in Ref. \cite{Lin2019}
that the conductance of the structure corresponds to a network with
one elastic strand removed. In this context it is worthwhile to mention
that the elastic contribution of loop strands $i=2$ measured in Ref.
\cite{Lange2011} is smaller by a factor of two as compared to the
average elastic contribution in networks made of stars with $f=4$.
This observation is consistent with the simulation data presented
in Figure 3 of Ref. \cite{Lang2018} and the prediction for this ratio
that is $\left(f-2\right)/f$ (compare equation (S29) of the supporting
information of Ref. \cite{Lang2018} with the contribution of a strand
within a perfect network). Ref. \cite{Lin2019} predicts at equation
(18) that strands $i=2$ contribute only a quarter of the ideal strands,
which is, therefore, neither in agreement with simulation data \cite{Lang2018}
nor with experiments \cite{Lange2011}.

Altogether, there is no principal disagreement between the ILGA and
older work \cite{Flory1982}, where it was discussed that the cycle
rank of a reduced network (where all pending structures connected
to the network only through a single point have been removed) is the
proper measure for the modulus of the network. This result is exact
only if the resistor analogy provides an exact representation of the
phantom modulus of an elastic network. However, this is not the case
when finite loops appear in the network as discussed in the following
section.

\section{\label{sec:Force-balance-approach}Force balance vs. resistor analogy}

To understand why the resistor analogy breaks down for polymer networks
containing finite cycles, we have to recall first that the free energy
of a Gaussian strand is a function of the square ratio of the size
$R$ of the strand with respect to its reference size \cite{Rubinstein2003},
$R_{0}=bN^{1/2}$. Size dependence cancels out for computing the entropy
of the chain and the degree of polymerization can be interpreted as
a resistance $\propto1/N$, when \emph{all strands are linked instantaneously}
with a size distribution identical to the unperturbed one at the reference
state. This is one essential step for deriving, for instance, the
affine model of rubber elasticity and it is the first condition that
needs to be satisfied to employ the resistor analogy. Note that this
condition is equivalent to considering individual elastic springs
that are all stretched by thermal agitation at the instance of cross-linking
and thus, are at equilibrium conformations.

The second condition is that a \emph{force balance} must be achieved
at each junction, since this refers to the maximum entropy of the
network (in the resistor analog, the sum of all currents at a node
must be zero \cite{DeGennes1976}). Finally, we have to sum over the
elastic contributions of \emph{all real strand parts} of the combined
chains that represent the elastic network to determine the phantom
modulus correctly \cite{James1949,Gusev2019}. This is the reason
why, for instance, $N/(N+N_{\text{eff}})$ appears in equation (\ref{eq:e2}).

\begin{figure}
\begin{centering}
\includegraphics[width=0.6\columnwidth]{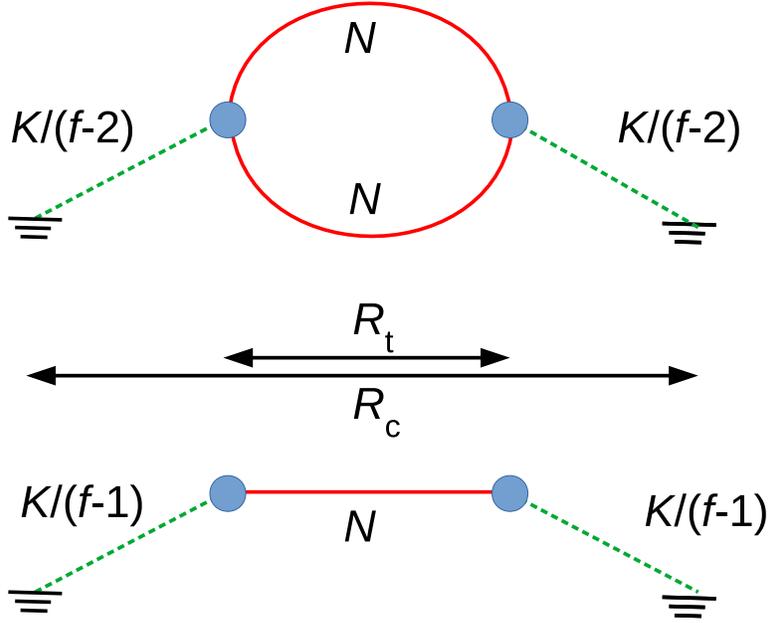}
\par\end{centering}
\caption{\label{fig:Comparison}Comparison of a loop made of two strands, $i=2$,
(top part) with a network strand in the ideal tree approximation (lower
part). The red lines indicates the real strands, the green dashed
lines are the virtual strands that connect to the non-fluctuating
elastic background (or to ground in the resistor analogy). $R_{\text{t}}$
and $R_{\text{c}}$ are the time average size of the real strands
and the combined strand respectively.}
\end{figure}

To elucidate the relationship between these conditions in case of
loop formation, let us consider the simplest non-pending loop $i=2$.
In Figure \ref{fig:Comparison}, this case is simplified to the linking
of two strands with $N$ segments into a loop while connecting the
loop by two strands of $K/(f-2)$ segments to ground. To satisfy the
first condition, we have to link instantaneously all strands at their
reference size. If we do so, the two ``parallel'' strands of the
loop $i=2$ pull with twice the force against the ``single'' strands
to ground, which does not refer to a force balance at the instance
of cross-linking. But when the structure relaxes forces, the strands
are no more at reference size. Thus, both conditions cannot be satisfied
at a time for this ``instantaneous linking'' (IL) case to form finite
loops.

Without loop formation as in the perfect tree approximation, there
are only linear connections of single strands, see Figure \ref{fig:Comparison},
which satisfies a force balance at the instance of cross-linking.
Only in this case, the ratio between the time average square size
of the real strand, $R_{\text{t}}^{2}$, and the square size of the
combined strand, $R_{\text{c}}^{2}$, is equivalent to the portion
of the combined chain that refers to the $N$-mer. Only then we can
replace the ratio of these square sizes by the degree of polymerization
of the real and the combined strand.

Now, let us consider that the loop is formed in reference size as
characterized by a single elastic strand of $N/2$ segments. Subsequent
linking of this relaxed loop to ground results then in a force balance
at the instance of cross-linking. Let us call this condition the ``relaxed
linking'' (RL) condition. However, we still have to consider here
that the loop contributes two real strands that are both stretched
to the time average size of the loop. This latter point is essentially
the difference between the results in Ref. \cite{Lin2019} for $i=2$
(where only one contribution is considered) and Refs. \cite{Zhong2016,Lang2018}
that consider two contributions.

The approach of Ref. \cite{Lang2018} up to equation (3) and sections
I-IV of the supporting information is considering RL, which assures
a force balance at the instance of cross-linking such that no further
relaxation of the strands must be considered. This latter point is
probably not obvious, since more complex loops are mapped onto the
case $i=2$ through a series of $Y-\Delta$ transforms \cite{Kenelly1899}.
The effect of these transforms is to split the elastic forces into
three disjoint contributions, that refer to a) the $N$-mer that connects
both adjacent junctions directly, b) an effective second strand that
connects these junctions not through ground (the ``bypass''), and
c) the sum of all contributions that run through ground. Bypass and
direct connection are combined into a single elastic strand 
\begin{equation}
N_{\text{X}}=\frac{N_{\text{k}}N}{N_{\text{k}}+N}\label{eq:NX}
\end{equation}
that is linked at its reference size with the strands to ground, which
refers to a simultaneous force balance at all junctions of the loop
(under consideration of all loop strands and all strands to ground)
for all $i\ge2$. A very recent paper by Panyukov \cite{Panyukov2019}
also uses this condition to establish a force balance at the loop
junctions. This procedure leads to a reduction in the elastic effectiveness
within the network as compared to a perfect tree network by

\begin{equation}
\Delta\epsilon_{2}\approx\frac{2(f-1)}{f^{2}}\label{eq:Deltae2}
\end{equation}
per strand of the loop for $i=2$ and by approximately 
\begin{equation}
\Delta\epsilon_{\text{i}}\approx\frac{2(f-2)}{(f-1)^{i}}\label{eq:Deltaei}
\end{equation}
per strand of the loop for $i\ge3$.

The opposite limit of IL was discussed in section VI of the supporting
information of Ref. \cite{Lang2018}. In comparison to RL, the elastic
contributions of a strand of the loop in case of IL, $\epsilon_{\text{i,IL}}$,
is larger than the contribution in case of RL,$\epsilon_{\text{i}}$,
according to 
\begin{equation}
\epsilon_{\text{i,IL}}=\epsilon_{\text{i}}\frac{N+2K_{\text{av}}}{N_{\text{X}}+2K_{\text{av}}}\label{eq:DEIL}
\end{equation}
for all $i\ge2$, since $N_{\text{X}}<N$. However, the $N_{\text{X}}$
of the RL case converges quickly towards $N$ that is considered for
IL, such that there is only a significant difference for the smallest
$i\apprge2$. Therefore, also the change in elastic effectiveness
per strand of the loop converges towards equation (\ref{eq:Deltaei})
for large $i$.

Ref. \cite{Lin2019} does not consider a force balance up to equation
(25) but uses the result from the resistor analogy that each pending
loops removes $kT$ of elastic energy plus an incorrect term for $i=2$
as discussed above. To remedy the missing force balance, the time
average square extensions of all loop sub-strands in the network are
assumed to be identical to the unperturbed loop sub-strands in sol
with a corresponding reduction of size and thus, an elastic contribution
$\propto1-1/i$. This correction refers to imposing a force balance
along the loop but ignores the virtual strands to ground. These actually
increase loop sizes as compared to the unperturbed size in sol, since
the bypass contains $>(i-1)N$ segments for $i>2$. This is also evident
from the simulation data in Figure 3 of Ref. \cite{Lang2018}, where
such a qualitative behavior of unperturbed loop strands can be observed
indeed, however,\emph{ only} \emph{after dividing out }the excess
strain of the network strands that enlarges loop sizes as compared
to isolated loops in sol. Altogether, the treatment of Ref. \cite{Lin2019}
overestimates the correction to modulus for all $i\ge2$.

Besides the differences between IL and Rl, we have to be aware that
these two limiting cases are a large simplification of the real cross-linking
process. It has been discussed in a series of recent works \cite{Guerin2012,Guerin2013,Guerin2014,Levernier2015,Dolgushev2015}
that loop formation within a single polymer is a non-Markovian process
that inevitably occurs at \emph{enlarged} chain conformations. The
discussion in these works provides strong arguments that loop formation
in networks should occur also at enlarged chain conformations. Since
cyclization competes with other reactions during network formation,
the results of Refs. \cite{Guerin2012,Guerin2013,Guerin2014,Levernier2015,Dolgushev2015}
are not quantitative concerning network formation. Besides these works,
other arguments were put forward in recent work to explain stretched
chain conformations in cross-linked networks \cite{Lang2017}. Altogether,
the assumption of equilibrium chain conformations at the instance
of cross-linking has to be questioned and further investigations are
necessary to understand under which conditions loops are closed and
elastic chains are incorporated into the network structure.

In any of the cases mentioned above, the chains are stretched at the
instance of cross-linking as compared to the chain conformations assumed
in the RL case. Thus, the consideration of RL should provide a lower
bound for the phantom modulus as discussed in the following section.

Finally, we have to remark that the resistor analogy can be used to
compute exactly the effective elastic strand that describes the fluctuations
of a given junction with respect to the position of an adjacent junction
or with respect to ground. This is relevant for analyzing the relaxation
dynamics of the network since the Kirchhoff matrix is the ``generalized
Rouse matrix'' of the network \cite{Gurtovenko}. But this connectivity
based analysis is not sufficient to characterize modulus, since phantom
modulus is a function of the time average the extension of the elastic
network strands. A related example for this point is the superelasticity
of de-swollen networks \cite{Obukhov1994}, where the de-swelling
causes shrunken chain conformations, and correspondingly, a lower
phantom modulus of the network.

\section{\label{sec:II.-The-distribution}A first estimate on the maximum
impact of finite loops on phantom modulus}

Several simulation studies addressed the average loop DP \cite{Michalk2001,Lang2007},
the distribution of loop DP's \cite{Schulz1992,Michalke2002,Lang2007},
or the frequency of smallest loops $\propto N^{-1/2}$ in end-linked
model networks \cite{Leung,Grest,Gilra,Schwenke2011,Lang2012a}. The
average loop DP has been estimated both analytically \cite{Cai2015}
and numerically \cite{Lang2007}. Furthermore, a recursion method
to compute the DP distribution of loops has been proposed \cite{Lang2001}
and it was shown that the loop DP distribution scales universally
$\propto i^{-5/2}$ at the gel point \cite{Kilb1958,Suematsu1998,Lang2005a,Wang2017}.
For well developed end-linked model networks far beyond the gel point,
there is a characteristic peak in the DP distribution \cite{Schulz1992,Michalke2002,Lang2007},
that is a function of conversion $p$, junction functionality $f$,
and the overlap number $P$ of elastic strands in the network \cite{Lang2007},
with an additional dependence on the stoichiometric imbalance $r$
in case of networks that were established by a co-polymerization,
since increasing stoichiometric imbalance drives the samples closer
to the gel point.

In comparison to these model networks, randomly cross-linked networks
have a different DP distribution of smallest loops that decays $\propto N^{-3/2}$
with only little dependence on conversion, since pending loops  result
essentially from the contacts between monomers of the same polymer
\cite{Lang2003}. While there are significant differences for the
smallest loops, the peak in the DP distribution and the average loop
DP agree almost quantitatively with the corresponding results for
the end-linked model networks \cite{Lang2007}. This latter point
was used in Ref. \cite{Panyukov2019} to estimate the characteristic
loop size by considering the corresponding end-linked model networks
instead of randomly cross-linked networks. But as we will see below,
the smallest loops contain the least elastically effective strands.
Thus, the estimate of Ref. \cite{Panyukov2019} for randomly cross-linked
networks is not quantitative for end-linked model networks.

For our simple first order estimate on the net impact of finite loops
on phantom modulus in end-linked model networks, we make use of the
result that the frequency $L_{\text{i}}$ of the smallest loops made
of a different number of strands, $i\ne j$, can be related \cite{Lang2005a,Schwenke2011}
through 
\begin{equation}
\frac{L_{\text{i}}}{L_{\text{j}}}\approx\left[p(f-1)\right]^{|i-j|}(j/i)^{5/2}.\label{eq:ratio2}
\end{equation}
This relation is obtained from adopting a mean-field estimate for
the average number of $\approx f\left[p(f-1)\right]^{i-1}$ connected
units in generation $i$ away within an incomplete Bethe lattice where
only a portion $p$ of bonds is existing. The number of connected
units provides the expected number of attempts to close a cycle containing
$i$ strands, while the probability for closure is related to the
return probability $\propto\left(iN\right)^{-3/2}$ of a random walk
containing $i$ strands. The power $5/2$ results from integrating
this return probability up to conversion $p$, see the discussion
following equation (20) in Ref. \cite{Lang2005a}.

Equation (\ref{eq:ratio2}) can be applied as long as the cycle rank,
$\xi$, is large compared to the frequency of the smallest loops,
$\xi\gg R_{1}$, since then, the formation of finite loops is only
a weak distortion of the incomplete Bethe lattice at small $i$. Note
that the above relation refers to the simplest case of a homo-polymerization
of $f$ functional star molecules and is based upon the assumption
that adjacent loops are formed independently from each other. More
complex architectures including end-linked model networks are discussed
elsewhere \cite{Suematsu2002,Lang2005a}.

Recall that in equation (\ref{eq:Deltaei}), the impact per loop strand
is $\propto\left(f-1\right)^{-i}$ for $i\ge3$. With this dependence,
we can relate the net effect of loops with a different number of elastic
strands $i>3$ in well developed networks within the above approximations:
\begin{equation}
\frac{i}{3}\frac{\Delta\epsilon_{\text{i}}L_{\text{i}}}{\Delta\epsilon_{3}L_{3}}\approx p^{i-3}\left(\frac{3}{i}\right)^{3/2}.\label{eq:i2}
\end{equation}
Here, the numerical coefficient of $i/3$ reflects that all $i$ strands
of a loop reduce modulus by $\Delta\epsilon_{i}$. Thus, the net impact
of loops $i\ge3$ can be written as 
\begin{equation}
\sum_{i=3}^{\infty}i\Delta\epsilon_{\text{i}}L_{\text{i}}\approx3^{5/2}\Delta\epsilon_{3}L_{3}\sum_{i=3}^{\infty}p^{i-3}i^{-3/2}.\label{eq:sum1}
\end{equation}
The quick decay as a function of $i$ in this equation shows that
still the largest net corrections to the ideal phantom modulus result
from the smallest $i$ despite the fact that much more strands are
part of large loops.

Simulation data on the smallest loops \cite{Lang2007} that can be
found for a given strand with a spanning tree approach \cite{Lang2001b}
indicate that there is a characteristic loop DP of $i_{c}N$ (a typical
example were $i_{c}\approx8$ for $N\approx20$ and $f=4$ at $p\approx1$,
see Ref. \cite{Lang2007}) at which the distribution of smallest loops
obtains a maximum. However, the strands that constitute such small
loops are also part of a huge number of larger loops that can be identified
within the network. We take equation (\ref{eq:ratio2}) as a mean-field
estimate for the abundance of these larger loops and thus, we do not
truncate the summation at the characteristic loop DP. Instead, we
use the infinite sum for a simplified mean-field estimate for the
net impact of all of these larger loops. Note that in the case of
full conversion, $p=1$, the infinite sum is at its maximum but remains
a number of order unity: $\sum_{i=3}^{\infty}i^{-3/2}\approx1.2588...\approx5/4$.

Altogether, we expect that the modulus $G$ close to full conversion,
$p\approx1$, is smaller than the ideal phantom modulus 
\begin{equation}
G_{\text{id}}=\xi kT/V=\frac{f-2}{f}\nu kT,\label{eq:Gid}
\end{equation}
where $\nu$ is the density of network strands, $k$ the Boltzmann
constant, $T$ the absolute temperature, and $V$ the volume of the
sample. A lower bound for modulus is estimated by considering the
corrections due to finite loops in the RL case 
\begin{equation}
G\approx G_{\text{id}}\left\{ 1-\frac{L_{1}}{\xi}-2\Delta\epsilon_{2}\frac{L_{2}}{\xi}-\sum_{i=3}^{\infty}i\Delta\epsilon_{\text{i}}\frac{L_{i}}{\xi}\right\} \label{eq:modulus estimate}
\end{equation}
\[
\approx G_{\text{id}}\left\{ 1-\frac{L_{1}}{\xi}\left(1+2^{-3/2}\left(f-1\right)\Delta\epsilon_{2}+\frac{5}{4}\left(f-1\right)^{2}\Delta\epsilon_{3}\right)\right\} 
\]
\[
\approx G_{\text{id}}\left\{ 1-\frac{L_{1}}{\xi}\left[1+\frac{\left(f-1\right)^{2}}{\sqrt{2}f^{2}}+\frac{5(f-2)}{2\left(f-1\right)}\right]\right\} 
\]
\[
=\frac{kT}{V}\left(\xi-c_{\text{f}}L_{1}\right).
\]
Above we introduced $c_{\text{f}}$ as short hand notation for the
term in the square brackets. Note that $c_{\text{f}}=7/2+2^{-1/2}$
for $f\rightarrow\infty$. For $f=3,$ there is $c_{\text{f}}\approx2.56$
and for $f=4$, there is $c_{\text{f}}\approx3.06$. Thus, the net
correction to modulus due to loops with $i\ge2$ is larger than the
correction due to pending loops $i=1$ for all $f\ge3$.

In practice, $L_{1}$ or $L_{2}$ can be determined through one of
the available methods \cite{Lange2011,Zhou2012} or by MALDI-TOF experiments
\cite{Kricheldorf2001} on an analog $f=2$ system, while $\xi$ can
be computed from the conversion of the reactive groups in the gel
\cite{Miller1976,Lang2004}. As pointed out in previous work \cite{Suematsu2002,Lang2005a,Wang2016},
we have to emphasize that cyclization is a unique function of the
overlap number $P\propto\phi N^{1/2}$ and the junction functionality.
Thus, at the absence of a quantitative method for counting loops,
one can still analyze qualitatively the corrections to modulus as
a function of the overlap number by recalling that \cite{Lang2005a}
\begin{equation}
L_{1}\propto\frac{f-1}{\phi N^{1/2}}\label{eq:R1}
\end{equation}
for the practically important limit of $L_{1}\ll\xi$ and small $f$
where multiple pending loops per junction can be neglected and where
the impact of finite loops is just a correction to the phantom modulus.

\begin{figure}
\includegraphics[angle=270,width=1\columnwidth]{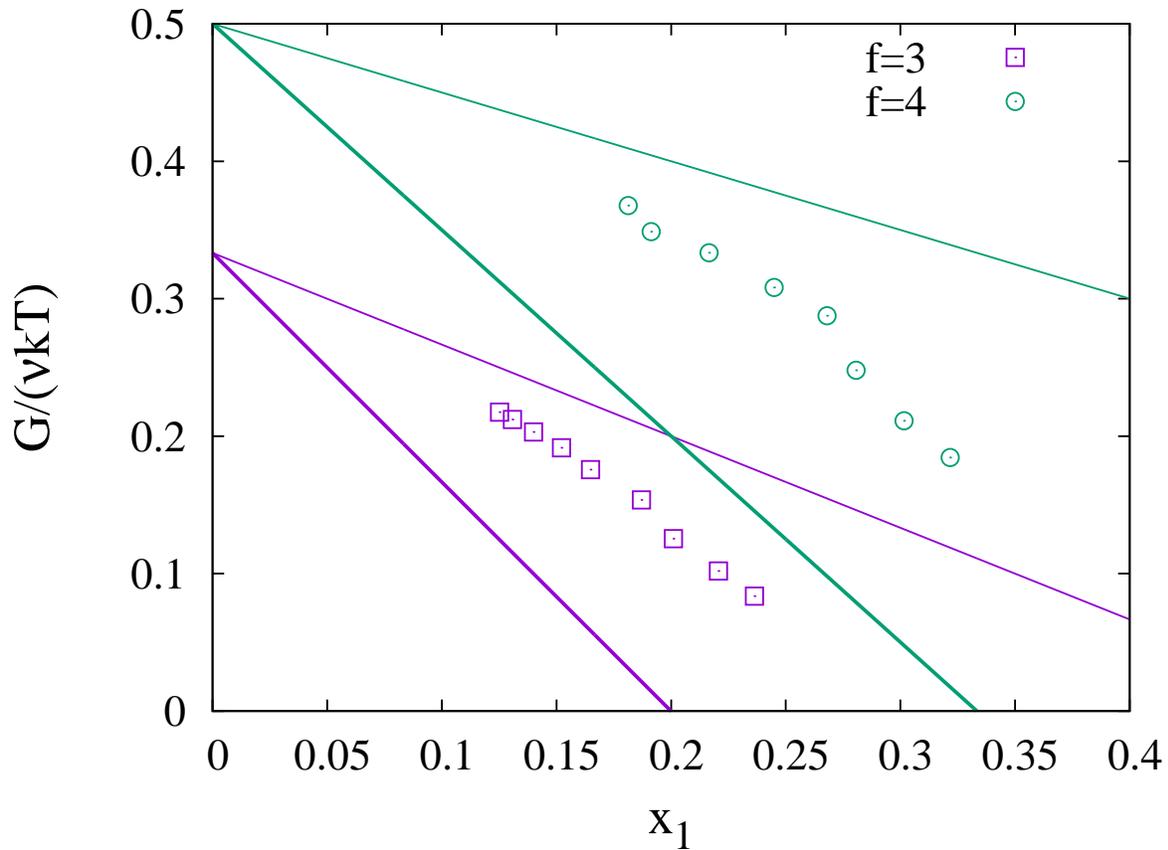}

\caption{\label{fig:Comparison-of-the}Comparison of the loop dependence of
modulus data of Ref. \cite{Lin2019} with the lower bound of equation
(\ref{eq:Gid1}) as shown by the thick lines, while the thin lines
indicate the corrections due to pending loops only, $c_{\text{f}}=1$,
which is equivalent to the estimate from the resistor analogy.}
\end{figure}

For a quantitative test of our discussion, we compare equation (\ref{eq:modulus estimate})
with the data of Refs. \cite{Zhong2016,Lin2019}. We assume $p=1$
as the effect of incomplete conversion is discussed in Ref \cite{Lin2019}
but no information on conversion is presented that would allow to
shift the experimental data accordingly. For infinite stoichiometric
end-linked model networks at $p=1$, there are $f/2-1$ loops per
network junction (ignoring loops in the sol). Thus, the number of
primary loops per junction, $x_{1}$, of Ref \cite{Lin2019} is related
to the fraction of primary loops among all loops through $L_{1}/\xi=2x_{1}/\left(f-2\right)$.
Thus, we expect that the experimental data should be above 
\begin{equation}
G_{\text{id}}\left(1-\frac{c_{\text{f}}L_{1}}{\xi}\right)\approx\left(f-2-2c_{\text{f}}x_{1}\right)\nu kT/f\label{eq:Gid1}
\end{equation}
and below the pending loop correction (which is the estimate resulting
from the resistor analogy) with $c_{\text{f}}=1$. Indeed these limits
set apparently an upper and a lower bound to the experimental data
of Refs. \cite{Zhong2016,Lin2019}, see Figure \ref{fig:Comparison-of-the}.

In Ref. \cite{Lin2019}, the gel point loop DP distribution $\propto i^{-5/2}$
was used in contrast to equation (\ref{eq:ratio2}), which largely
underestimates the frequency of loops $i\ge2$, while the impact of
finite loops was overestimated for $i\ge2$ as discussed in the preceding
section. The resulting agreement with experimental data in Ref. \cite{Lin2019}
is, therefore, less the indication of a precise approximation than
merely the fortuitous cancellation of the following corrections that
we have not yet taken into account.

First, the above treatment is based upon the impact of a single loop
in an otherwise perfect network, which is a rather rough approximation,
since essentially the whole network consists of loops of similar DP
as the one that is considered. In a network of loops, the junction
fluctuations will be further enhanced as the junctions are connected
to less neighboring junctions in the further environment. Larger junction
fluctuations imply a smaller elastic contribution as the real part
of the corresponding combined strands is being reduced.

Second, with increasing conversion, it is expected that reactions
involve junctions that are increasingly apart from each other. This
holds in particular when the available number of reactive groups drops
below one per pervaded volume of a network strand. Entanglements provide
a bias towards enlarged chain conformations at cross-linking as argued
in Ref. \cite{Lang2017} and loop closure in general occurs at enlarged
chain conformations \cite{Guerin2012,Guerin2013,Guerin2014,Levernier2015,Dolgushev2015}.
Furthermore, the IL case discussed above occurs at enlarged chain
conformations as compared to the RL limit. As all these cases lead
to larger contributions to modulus as the estimate in equation \ref{eq:Gid1},
the latter must be a lower limit for modulus, given that all other
corrections (besides enlarged chain conformations at loop closure)
were treated correctly.

Third, the contribution of entanglements to modulus is ignored in
the above discussion. This contribution is a function of the weight
fraction of elastically active polymer at cross-linking conditions,
$\omega$, and scales $\propto\omega^{7/3}$ as discussed in Refs.
\cite{Adam1983,Rubinstein2003}. This contribution cannot be ignored
by arguing that the precursor polymers were not entangled, since the
active network strands are parts of ``infinitely long chains'' that
are always above a finite concentration-dependent entanglement degree
of polymerization. Since the data in Ref. \cite{Zhong2016} was obtained
by diluting the reaction mixture, there is an additional small entanglement
contribution to modulus hidden in Figure \ref{fig:Comparison-of-the},
which scales roughly $\propto x_{1}^{-7/3}$, if $\omega\approx\phi$
and $x_{1}\sim\phi^{-1}$.

Finally, the original analysis of primary loops indicated a significant
amount of pending chains, see Figure 4c of Ref. \cite{Zhou2012}.
In subsequent work by the same group \cite{Zhong2016,Lin2019}, the
corresponding conversion $p<1$ was not provided. The measured modulus,
therefore, must lie below the ideal prediction, which is not yet accounted
for in the analysis.

Altogether, the ILGA is probably a reasonable first step towards the
development of more advanced models for phantom modulus. But it is
not the only correction that is relevant for an improved estimate
of the phantom modulus. One key problem for further development of
theory is a self-consistent representation of the surrounding network
structure, which is addressed to some extent in the following section.

\section{\label{sec:Polydisperse-network-strands}Some remarks on imperfect
and polydisperse networks}

One of the largest simplifications in the computations of Refs. \cite{Zhong2016,Lang2018,Lin2019}
is that the surrounding network is represented by the very same elastic
strand to ground at any junction. In real networks, there are pending
chains, loops of different $i$, polydisperse strands, etc ... that
may cause a broad distribution of elastic strands to ground and it
is not a priori clear that this distribution can be represented by
a single (average) effective strand. Therefore, we consider in the
present section the effect of a distribution of elastic strands on
elasticity and junction fluctuations.

As the simplest possible example, let us consider a perfect tree network
made of $f$-functional junctions that are connected by a polydisperse
distribution of chains that is described by a number fraction distribution
$n_{\text{N}}$. It has been shown previously \cite{Graessley1975,Higgs1988},
that one recovers the classical phantom coefficient of $\left(f-2\right)/f$
also in the case of polydisperse networks within the resistor analogy
and the ideal tree approximation. The author of the present work has
repeated the numerical calculations of Ref. \cite{Higgs1988} for
several model distributions and can confirm that this result is correct
independent of the shape of the distribution. But as shown for the
numerical example below, average quantities (except of modulus) like
junction fluctuations, elastic strands to ground, or network strands
between junctions are not representative of the behavior of the full
distributions.

As a representative numerical example, let us consider a most probable
distribution of elastic strands between the junctions that is characterized
by a number fraction distribution of 
\begin{equation}
n_{\text{N}}(p)=(1-p)p^{N-1},\label{eq:most probable}
\end{equation}
where $p$ is the probability that a monomer is not cross-linked.
The corresponding number average degree of polymerization is $N_{\text{n}}=\left(1-p\right)^{-1}$
and the weight fractions of $N$-mers are here 
\begin{equation}
w_{\text{N}}(p)=N(1-p)^{2}p^{N-1}.\label{eq:wN}
\end{equation}

\begin{figure}
\includegraphics[angle=270,width=1\columnwidth]{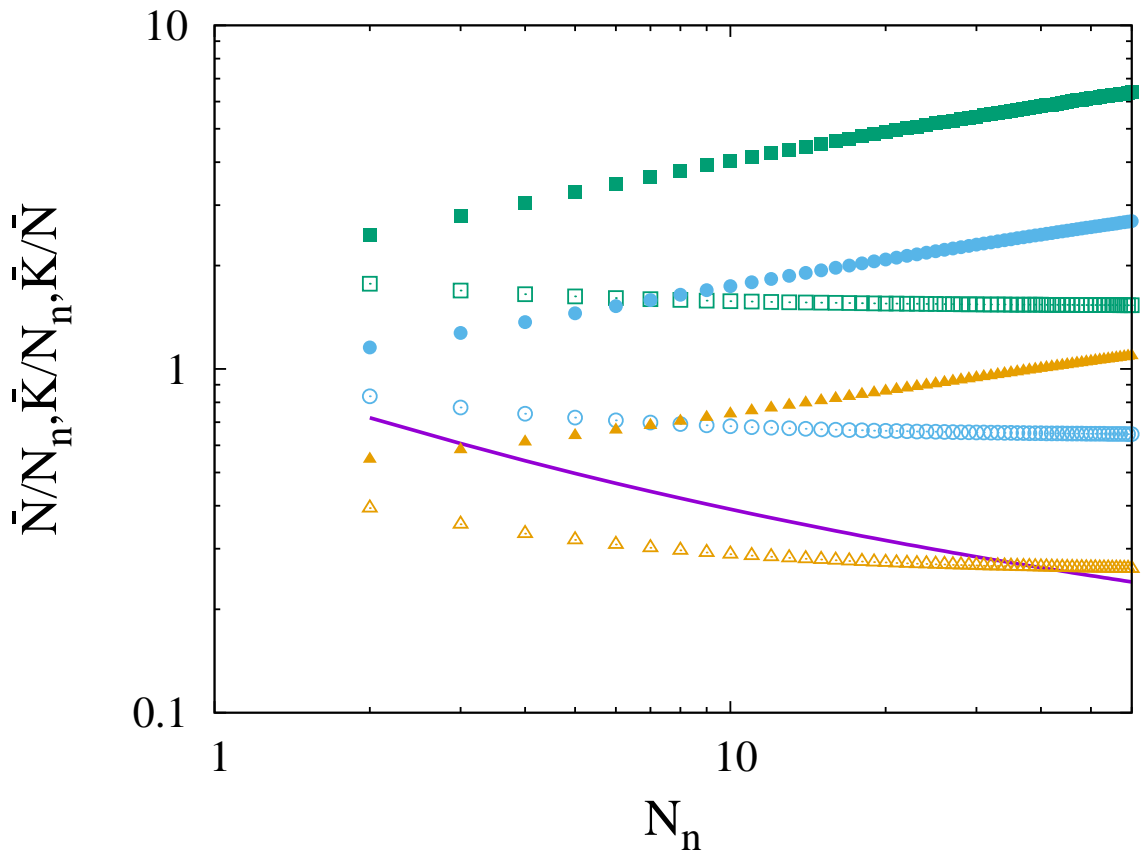}

\caption{\label{fig:Comparison-of-average}Comparison of average conductances
$1/\overline{N}$, the number average degree of polymerization $N_{\text{n}}$
and the average conductance to ground, $1/\overline{K}$ for perfect
ideal tree network with a most probable weight distribution of strands.
The line is $\overline{N}/N_{\text{n}}$, the squares, circles, and
triangles refer to $f=$ 3, 4, 6 respectively. Open symbols are $\overline{K}/N_{\text{n}}$,
while closed symbols refer to $\overline{K}/\overline{N}$. }
\end{figure}

Recall from equation (\ref{eq:Ne}) that conductances are required
to compute the elastic contribution of network strands. Equation (\ref{eq:most probable})
leads for this model case to an ensemble average ``conductance'',
$1/\overline{N}$ of the polymers between two junctions, for which
a simple analytical expression is available 
\begin{equation}
\frac{1}{\overline{N}}=(1-p)\sum_{N>0}\frac{p^{N-1}}{N}=\frac{\left(1-p\right)\ln\left(1-p\right)}{-p}=\frac{\ln\left(N_{\text{n}}\right)}{N_{\text{n}}-1}.\label{eq:Nr poly}
\end{equation}
Note that $1/\overline{N}$ is not $\propto N_{\text{n}}^{-1}$ as
the phantom modulus. Similarly, one finds with the self-consistent
numerical approach introduced in Ref. \cite{Higgs1988} that also
the average conductance to ground, $1/\overline{K}$, is not a simple
function of neither $1/\overline{N}$ nor $N_{\text{n}}^{-1}$. This
is illustrated in Figure \ref{fig:Comparison-of-average} that shows
several ratios between $\overline{K}$, $\overline{N}$, and $N_{\text{n}}$,
which all are not constant as a function of the cross-link density.
Therefore, the average $\overline{K}$ and $\overline{N}$ do not
compensate each other for producing a constant phantom coefficient
of $(f-2)/f$. Compensation is only achieved when the full distributions
for $\overline{K}$ and $\overline{N}$ are taken into account.

\begin{figure}
\includegraphics[angle=270,width=1\columnwidth]{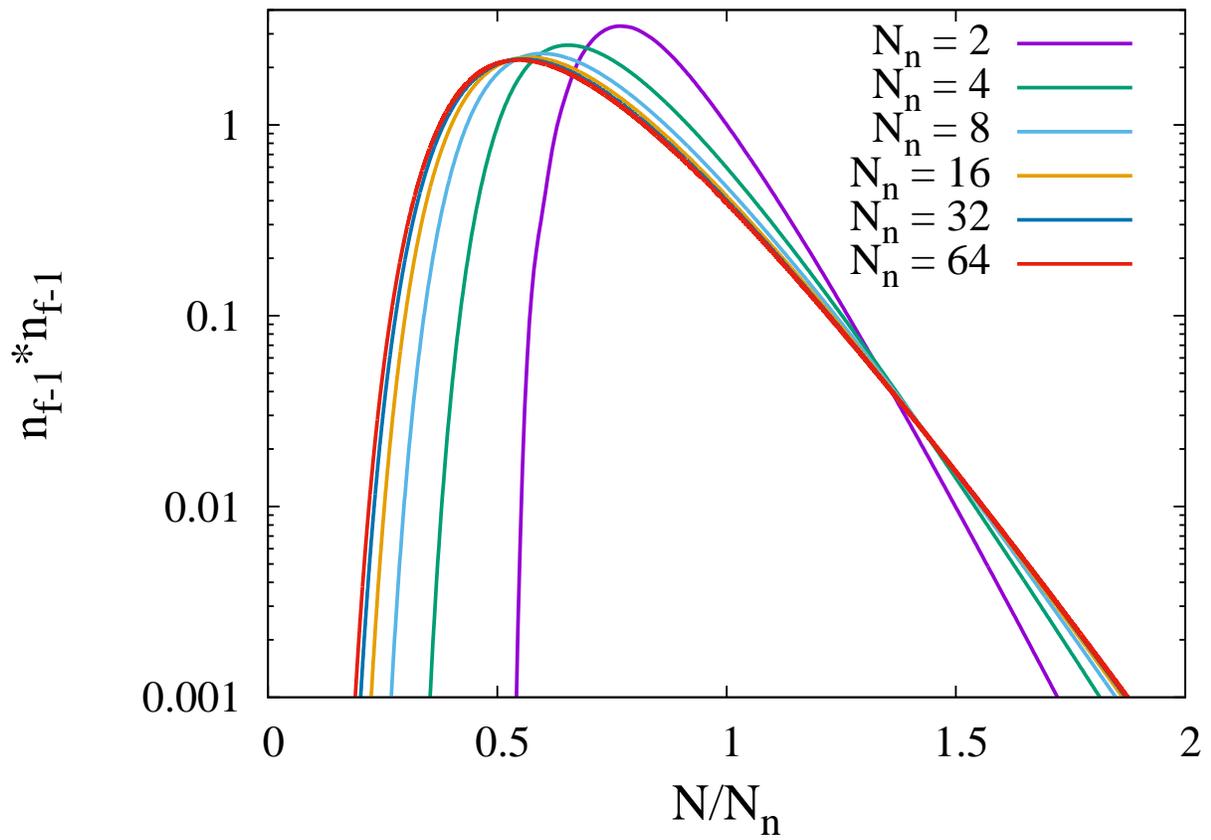}

\caption{\label{fig:Distribution-of-the}Distribution of the virtual strands
to ground, $n_{\text{f-1}}*n_{\text{f-1}}$, for different $N_{\text{n}}$. }
\end{figure}

The reason for this point is that the distributions of $K$ drift
and change shape as a function of $N_{\text{n}}$. In consequence,
also the net distribution $n_{\text{f-1}}$ of the $f-1$ virtual
strands to ground at one junction changes shape as a function of $N_{\text{n}}$.
This is shown in Figure \ref{fig:Distribution-of-the} for the convolution
of the virtual strand distributions at both junctions, $n_{\text{f-1}}*n_{\text{f-1}}$.
Relevant for elasticity is the distribution of the combined strands
$n_{\text{c}}$ that is given through yet another convolution with
the distribution of network strands, $n_{\text{N}}$: 
\begin{equation}
n_{\text{c}}=n_{\text{f-1}}*n_{\text{f-1}}*n_{\text{N}},\label{eq:convolution}
\end{equation}
Note that in Refs. \cite{Lang2010,Lang2013} it was discussed that
the ensemble average vector order parameter in a model network is
the inverse of the degree of polymerization of the corresponding combined
chain. Thus, we expect that the distribution of the inverse $n_{\text{c}}$
values should be relevant to characterize a polydisperse or imperfect
network.

Let us denote by $n_{\text{c}}(N)$ the distribution of combined strands
that is obtained for a given $N$-mer in the middle instead of considering
the full distribution $n_{\text{N}}$. With this notation and the
results of preceding work \cite{Graessley1975,Higgs1988}, we can
write for the coefficient of the phantom model of a polydisperse perfect
tree network that 
\begin{equation}
\frac{G_{\text{ph}}}{\nu kT}=\sum_{N=1}^{\infty}n_{\text{N}}\sum_{n_{\text{c}}(N)}\frac{N}{n_{\text{c}}(N)}\label{eq:summation}
\end{equation}
\[
=N_{\text{n}}\sum_{N=1}^{\infty}w_{\text{N}}\sum_{n_{\text{c}}(N)}\frac{1}{n_{\text{c}}\left(N\right)}=\left\langle m\right\rangle N_{\text{n}}=\frac{f-2}{f}.
\]
Here, $\left\langle m\right\rangle $ is the ensemble average vector
order parameter (the ensemble average residual bond orientation in
the limit of infinitely long times, see Refs. \cite{Lang2010,Lang2013}
for more details). We see from equation (\ref{eq:summation}) that
the computation of the ensemble average elastic contribution of the
network strands refers to a weight average where the real chain part
of the combined elastic strands serves as a weight for the inverse
distribution of combined strands, $n_{\text{c}}(N)^{-1}$. The first
expression in second line of equation (\ref{eq:summation}) (without
$N_{\text{n}}$) is the summation rule for taking the ensemble average
residual bond orientation. Thus, the multi quantum nuclear magnetic
resonance (MQ-NMR) as described in Refs. \cite{CohenAddad,Saalwaechter}
is the method of choice to detect the distribution of elastic strands
in networks where the phantom model provides a reasonable approximation
for the shear modulus of the network. Further development of the theory
of phantom modulus can use this connection for a test of model predictions
against experimental data.

\section{Summary}

It was shown using the resistor analogy that only pending cycles within
an otherwise ``perfect tree'' network of monodisperse strands affect
the average elastic strand between adjacent junctions in the network.
Therefore, one arrives within the resistor analogy at the result that
the modulus of such idealized model networks should be fully determined
by the cycle rank after removing pending cycles \cite{Flory1982}.
It was further confirmed numerically that polydispersity has no effect
on elasticity, if the resistor analogy can be applied in agreement
with earlier work \cite{Graessley1975,Higgs1988}. Both elasticity
and the residual bond orientation as accessible through NMR, result
from a weight average of the inverse of the distribution of combined
elastic strands (containing the real polymer and the virtual strands
to ground). Thus, NMR is the method of choice to analyze the distribution
of elastic strands, in particular in imperfect or poly-disperse networks.
This is of interest also for networks made of finite loops, as the
most important non-pending finite loop $i=2$ can be separated for
appropriate parameters of the experiments \cite{Lange2011}.

The resistor analogy is no more exact once finite loops appear in
a network, since it is not possible to link the strands of the loop
at equilibrium conformations (of individual strands) while maintaining
simultaneously a force balance at the junctions. Assuming loops with
equilibrium conformations and a force balance at all loop junctions,
a lower bound estimate for the phantom modulus, $G_{\text{ph}}\approx\left(\xi-c_{\text{f}}L_{1}\right)kT/V$
is obtained, where $L_{1}$ is the number of primary (``pending'')
loops among the number of independent loops (the cycle rank $\xi$)
of the network, $k$ the Boltzmann constant, and $T$ the absolute
temperature. $c_{\text{f}}$ is a functionality dependent coefficient
that is $\approx2.56$ for $f=3$ and $\approx3.06$ for $f=4$, while
it converges quickly towards $\approx4.2$ in the limit of large $f$.
The resistor analogy provides an upper bound for modulus with $c_{\text{f}}=1$
that reflects the impact of pending loops only.

Further model development is necessary to understand loop closure
inside a network and the corresponding enlarged loop conformations.
Furthermore, a realistic self-consistent model for the surrounding
network made of finite cycles, containing defects, and built of polydisperse
strands must be developed. The recent paper by Panyukov \cite{Panyukov2019}
is a big step forward in this direction. The discussion in the present
work provides some guidelines for the construction of a successful
model.

\section{Associated Content}

Supporting Information
The supporting information (PDF) contains a brief FORTRAN code that provides 
a numerical proof for the resistor analogy within the ideal loop
gas approximation (ILGA).

\section{Acknowledgements}

This work was supported by the Deutsche Forschungsgemeinschaft (DFG) throgh Grants LA2735/5-1 and LA2735/6-1.

\providecommand{\latin}[1]{#1}
\providecommand*\mcitethebibliography{\thebibliography}
\csname @ifundefined\endcsname{endmcitethebibliography}
  {\let\endmcitethebibliography\endthebibliography}{}

\newpage
\begin{figure}
\includegraphics[width=1.0\columnwidth]{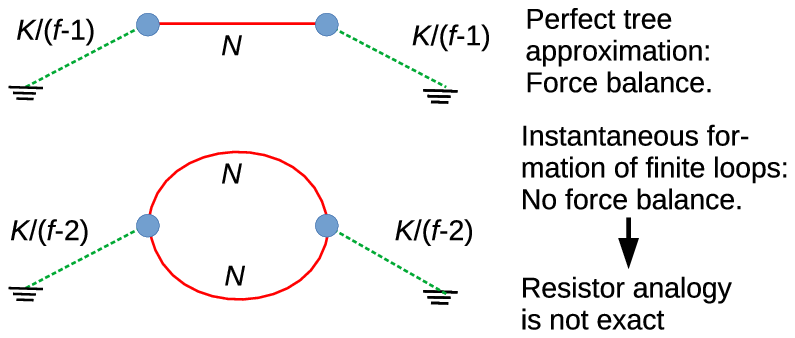}
\end{figure}


\begin{mcitethebibliography}{51}
\providecommand*\natexlab[1]{#1}
\providecommand*\mciteSetBstSublistMode[1]{}
\providecommand*\mciteSetBstMaxWidthForm[2]{}
\providecommand*\mciteBstWouldAddEndPuncttrue
  {\def\EndOfBibitem{\unskip.}}
\providecommand*\mciteBstWouldAddEndPunctfalse
  {\let\EndOfBibitem\relax}
\providecommand*\mciteSetBstMidEndSepPunct[3]{}
\providecommand*\mciteSetBstSublistLabelBeginEnd[3]{}
\providecommand*\EndOfBibitem{}
\mciteSetBstSublistMode{f}
\mciteSetBstMaxWidthForm{subitem}{(\alph{mcitesubitemcount})}
\mciteSetBstSublistLabelBeginEnd
  {\mcitemaxwidthsubitemform\space}
  {\relax}
  {\relax}

\bibitem[Lange \latin{et~al.}(2011)Lange, Schwenke, Kurakazu, Akagi, Chung,
  Lang, Sommer, Sakai, and Saalw{\"{a}}chter]{Lange2011}
Lange,~F.; Schwenke,~K.; Kurakazu,~M.; Akagi,~Y.; Chung,~U.~I.; Lang,~M.;
  Sommer,~J.-U.; Sakai,~T.; Saalw{\"{a}}chter,~K. {Connectivity and structural
  defects in model hydrogels: A combined proton NMR and Monte Carlo simulation
  study}. \emph{Macromolecules} \textbf{2011}, \emph{44}, 9666--9674\relax
\mciteBstWouldAddEndPuncttrue
\mciteSetBstMidEndSepPunct{\mcitedefaultmidpunct}
{\mcitedefaultendpunct}{\mcitedefaultseppunct}\relax
\EndOfBibitem
\bibitem[Zhou \latin{et~al.}(2012)Zhou, Woo, Cok, Wang, Olsen, and
  Johnson]{Zhou2012}
Zhou,~H.; Woo,~J.; Cok,~A.~M.; Wang,~M.; Olsen,~B.~D.; Johnson,~J.~A. Counting
  primary loops in polymer gels. \emph{PNAS} \textbf{2012}, \emph{109},
  19119--19124\relax
\mciteBstWouldAddEndPuncttrue
\mciteSetBstMidEndSepPunct{\mcitedefaultmidpunct}
{\mcitedefaultendpunct}{\mcitedefaultseppunct}\relax
\EndOfBibitem
\bibitem[Zhong \latin{et~al.}(2016)Zhong, Wang, Kawamoto, Olsen, and
  Johnson]{Zhong2016}
Zhong,~M.; Wang,~R.; Kawamoto,~K.; Olsen,~B.~D.; Johnson,~J.~A. Quantifying the
  impact of molecular defects on polymer network elasticity. \emph{Science}
  \textbf{2016}, \emph{353}, 1264--1268\relax
\mciteBstWouldAddEndPuncttrue
\mciteSetBstMidEndSepPunct{\mcitedefaultmidpunct}
{\mcitedefaultendpunct}{\mcitedefaultseppunct}\relax
\EndOfBibitem
\bibitem[Lang(2018)]{Lang2018}
Lang,~M. On the elasticity of phantom model networks with cyclic defects.
  \emph{ACS Macro Letters} \textbf{2018}, \emph{7}, 536--539\relax
\mciteBstWouldAddEndPuncttrue
\mciteSetBstMidEndSepPunct{\mcitedefaultmidpunct}
{\mcitedefaultendpunct}{\mcitedefaultseppunct}\relax
\EndOfBibitem
\bibitem[Lin \latin{et~al.}(2019)Lin, Wang, Johnson, and Olsen]{Lin2019}
Lin,~T.-S.; Wang,~R.; Johnson,~J.~A.; Olsen,~B.~D. Revisiting the Elasticity
  Theory for Real Gaussian Phantom Networks. \emph{Macromolecules}
  \textbf{2019}, \emph{52}, 1685--1694\relax
\mciteBstWouldAddEndPuncttrue
\mciteSetBstMidEndSepPunct{\mcitedefaultmidpunct}
{\mcitedefaultendpunct}{\mcitedefaultseppunct}\relax
\EndOfBibitem
\bibitem[Gusev(2019)]{Gusev2019}
Gusev,~A.~A. Numerical Estimates of the Topological Effects in the Elasticity
  of Gaussian Polymer Networks and Their Exact Theoretical Description.
  \emph{Macromolecules} \textbf{2019}, \emph{52}, 3244--3251\relax
\mciteBstWouldAddEndPuncttrue
\mciteSetBstMidEndSepPunct{\mcitedefaultmidpunct}
{\mcitedefaultendpunct}{\mcitedefaultseppunct}\relax
\EndOfBibitem
\bibitem[Flory(1982)]{Flory1982}
Flory,~P.~J. {Elastic activity of imperfect networks}. \emph{Macromolecules}
  \textbf{1982}, \emph{15}, 99--100\relax
\mciteBstWouldAddEndPuncttrue
\mciteSetBstMidEndSepPunct{\mcitedefaultmidpunct}
{\mcitedefaultendpunct}{\mcitedefaultseppunct}\relax
\EndOfBibitem
\bibitem[Graessley(1975)]{Graessley1975}
Graessley,~W.~W. {Elasticity and Chain Dimensions in Gaussian Networks}.
  \emph{Macromolecules} \textbf{1975}, \emph{8}, 865--868\relax
\mciteBstWouldAddEndPuncttrue
\mciteSetBstMidEndSepPunct{\mcitedefaultmidpunct}
{\mcitedefaultendpunct}{\mcitedefaultseppunct}\relax
\EndOfBibitem
\bibitem[Flory \latin{et~al.}(1976)Flory, Gordon, and McCrum]{Flory1976}
Flory,~P.; Gordon,~M.; McCrum,~N. {Statistical thermodynamics of random
  networks}. \emph{Proceedings of the Royal Society of London. Series A,
  Mathematical and Physical Sciences} \textbf{1976}, \emph{351}, 351--380\relax
\mciteBstWouldAddEndPuncttrue
\mciteSetBstMidEndSepPunct{\mcitedefaultmidpunct}
{\mcitedefaultendpunct}{\mcitedefaultseppunct}\relax
\EndOfBibitem
\bibitem[Higgs and Ball(1988)Higgs, and Ball]{Higgs1988}
Higgs,~P.; Ball,~R. {Polydisperse polymer networks : elasticity, orientational
  properties, and small angle neutron scattering}. \emph{Journal de Physique}
  \textbf{1988}, \emph{49}, 1785--1811\relax
\mciteBstWouldAddEndPuncttrue
\mciteSetBstMidEndSepPunct{\mcitedefaultmidpunct}
{\mcitedefaultendpunct}{\mcitedefaultseppunct}\relax
\EndOfBibitem
\bibitem[Panyukov(2019)]{Panyukov2019}
Panyukov,~S.~P. Loops in polymer networks. \emph{Macromolecules} \textbf{2019},
  \emph{52}, 4145--4153\relax
\mciteBstWouldAddEndPuncttrue
\mciteSetBstMidEndSepPunct{\mcitedefaultmidpunct}
{\mcitedefaultendpunct}{\mcitedefaultseppunct}\relax
\EndOfBibitem
\bibitem[Rubinstein and Colby(2003)Rubinstein, and Colby]{Rubinstein2003}
Rubinstein,~M.; Colby,~R.~H. \emph{{Polymer Physics}}; Oxford University Press,
  2003\relax
\mciteBstWouldAddEndPuncttrue
\mciteSetBstMidEndSepPunct{\mcitedefaultmidpunct}
{\mcitedefaultendpunct}{\mcitedefaultseppunct}\relax
\EndOfBibitem
\bibitem[De~Gennes(1976)]{DeGennes1976}
De~Gennes,~P.~G. On a relation between percolation theory and the elasticity of
  gels. \emph{Journal de Physique Lettres} \textbf{1976}, \emph{37},
  L1--L2\relax
\mciteBstWouldAddEndPuncttrue
\mciteSetBstMidEndSepPunct{\mcitedefaultmidpunct}
{\mcitedefaultendpunct}{\mcitedefaultseppunct}\relax
\EndOfBibitem
\bibitem[James and Guth(1949)James, and Guth]{James1949}
James,~H.~M.; Guth,~E. Simple Presentation of Network Theory of Rubber, with a
  Discussion of Other Theories. \emph{J. Polym. Sci.} \textbf{1949}, \emph{4},
  153--182\relax
\mciteBstWouldAddEndPuncttrue
\mciteSetBstMidEndSepPunct{\mcitedefaultmidpunct}
{\mcitedefaultendpunct}{\mcitedefaultseppunct}\relax
\EndOfBibitem
\bibitem[Kenelly(1899)]{Kenelly1899}
Kenelly,~A.~E. Equivalence of triangles and three-pointed stars in conducting
  networks. \emph{Electrical World and Engineer} \textbf{1899}, \emph{34},
  413--414\relax
\mciteBstWouldAddEndPuncttrue
\mciteSetBstMidEndSepPunct{\mcitedefaultmidpunct}
{\mcitedefaultendpunct}{\mcitedefaultseppunct}\relax
\EndOfBibitem
\bibitem[Guerin \latin{et~al.}(2012)Guerin, Benichou, and
  Voituriez]{Guerin2012}
Guerin,~T.; Benichou,~O.; Voituriez,~R. Non-Markovian polymer reaction
  kinetics. \emph{Nature Chemistry} \textbf{2012}, \emph{4}, 568--573\relax
\mciteBstWouldAddEndPuncttrue
\mciteSetBstMidEndSepPunct{\mcitedefaultmidpunct}
{\mcitedefaultendpunct}{\mcitedefaultseppunct}\relax
\EndOfBibitem
\bibitem[Guerin \latin{et~al.}(2013)Guerin, Benichou, and
  Voituriez]{Guerin2013}
Guerin,~T.; Benichou,~O.; Voituriez,~R. Reactive conformations and
  non-Markovian cyclization kinetics of a Rouse polymer. \emph{J. Chem. Phys.}
  \textbf{2013}, \emph{138}, 094908\relax
\mciteBstWouldAddEndPuncttrue
\mciteSetBstMidEndSepPunct{\mcitedefaultmidpunct}
{\mcitedefaultendpunct}{\mcitedefaultseppunct}\relax
\EndOfBibitem
\bibitem[Guerin \latin{et~al.}(2014)Guerin, Dolgushev, Benichou, Voituriez, and
  Blumen]{Guerin2014}
Guerin,~T.; Dolgushev,~M.; Benichou,~O.; Voituriez,~R.; Blumen,~A. Cyclization
  kinetics of Gaussian semiflexible polymer chains. \emph{Phys. Rev. E}
  \textbf{2014}, \emph{90}, 052601\relax
\mciteBstWouldAddEndPuncttrue
\mciteSetBstMidEndSepPunct{\mcitedefaultmidpunct}
{\mcitedefaultendpunct}{\mcitedefaultseppunct}\relax
\EndOfBibitem
\bibitem[Levernier \latin{et~al.}(2015)Levernier, Dolgushev, Benichou, Blumen,
  Guerin, and Voituriez]{Levernier2015}
Levernier,~N.; Dolgushev,~M.; Benichou,~O.; Blumen,~A.; Guerin,~T.;
  Voituriez,~R. Non-Markovian closure kinetics of flexible polymers with
  hydrodynamic interactions. \emph{J. Chem. Phys.} \textbf{2015}, \emph{143},
  204108\relax
\mciteBstWouldAddEndPuncttrue
\mciteSetBstMidEndSepPunct{\mcitedefaultmidpunct}
{\mcitedefaultendpunct}{\mcitedefaultseppunct}\relax
\EndOfBibitem
\bibitem[Dolgushev \latin{et~al.}(2015)Dolgushev, Guerin, Blumen, Benichou, and
  Voituriez]{Dolgushev2015}
Dolgushev,~M.; Guerin,~T.; Blumen,~A.; Benichou,~O.; Voituriez,~R. Contact
  Kinetics in Fracatal Macromolecules. \emph{Phys. Rev. Lett} \textbf{2015},
  \emph{115}, 208301\relax
\mciteBstWouldAddEndPuncttrue
\mciteSetBstMidEndSepPunct{\mcitedefaultmidpunct}
{\mcitedefaultendpunct}{\mcitedefaultseppunct}\relax
\EndOfBibitem
\bibitem[Lang(2017)]{Lang2017}
Lang,~M. Relation between Cross-Link Fluctuations and Elasticity in Entangled
  Polymer Networks. \emph{Macromolecules} \textbf{2017}, \emph{50},
  2547--2555\relax
\mciteBstWouldAddEndPuncttrue
\mciteSetBstMidEndSepPunct{\mcitedefaultmidpunct}
{\mcitedefaultendpunct}{\mcitedefaultseppunct}\relax
\EndOfBibitem
\bibitem[Gurtovenko and Blumen(2005)Gurtovenko, and Blumen]{Gurtovenko}
Gurtovenko,~A.~A.; Blumen,~A. Generalized Gaussian Structures: Models for
  Polymer Systems with Complex Topologies. \emph{Adv. Polym. Sci.}
  \textbf{2005}, \emph{182}, 171--282\relax
\mciteBstWouldAddEndPuncttrue
\mciteSetBstMidEndSepPunct{\mcitedefaultmidpunct}
{\mcitedefaultendpunct}{\mcitedefaultseppunct}\relax
\EndOfBibitem
\bibitem[Obukhov \latin{et~al.}(1994)Obukhov, Rubinstein, and
  Colby]{Obukhov1994}
Obukhov,~S.~P.; Rubinstein,~M.; Colby,~R.~H. Network Modulus and
  Superelasticity. \emph{Macromolecules} \textbf{1994}, \emph{27},
  3191--3198\relax
\mciteBstWouldAddEndPuncttrue
\mciteSetBstMidEndSepPunct{\mcitedefaultmidpunct}
{\mcitedefaultendpunct}{\mcitedefaultseppunct}\relax
\EndOfBibitem
\bibitem[Michalke \latin{et~al.}(2001)Michalke, Lang, Kreitmeier, and
  G{\"o}ritz]{Michalk2001}
Michalke,~W.; Lang,~M.; Kreitmeier,~S.; G{\"o}ritz,~D. Simulations on the
  number of entanglements of a polymer network using knot theory. \emph{Phys.
  Rev. E} \textbf{2001}, \emph{64}, 012801\relax
\mciteBstWouldAddEndPuncttrue
\mciteSetBstMidEndSepPunct{\mcitedefaultmidpunct}
{\mcitedefaultendpunct}{\mcitedefaultseppunct}\relax
\EndOfBibitem
\bibitem[Lang \latin{et~al.}(2007)Lang, Kreitmeier, and G{\"{o}}ritz]{Lang2007}
Lang,~M.; Kreitmeier,~S.; G{\"{o}}ritz,~D. {Trapped Entanglements in Polymer
  Networks}. \emph{Rubber Chemistry and Technology} \textbf{2007}, \emph{80},
  873--894\relax
\mciteBstWouldAddEndPuncttrue
\mciteSetBstMidEndSepPunct{\mcitedefaultmidpunct}
{\mcitedefaultendpunct}{\mcitedefaultseppunct}\relax
\EndOfBibitem
\bibitem[Schulz and Sommer(1992)Schulz, and Sommer]{Schulz1992}
Schulz,~M.; Sommer,~J.-U. Monte Carlo studies of polymer network formation.
  \emph{The Journal of Chemical Physics} \textbf{1992}, \emph{96},
  7102--7107\relax
\mciteBstWouldAddEndPuncttrue
\mciteSetBstMidEndSepPunct{\mcitedefaultmidpunct}
{\mcitedefaultendpunct}{\mcitedefaultseppunct}\relax
\EndOfBibitem
\bibitem[Michalke \latin{et~al.}(2002)Michalke, Kreitmeier, and
  G{\"{o}}ritz]{Michalke2002}
Michalke,~W.; Kreitmeier,~S.; G{\"{o}}ritz,~D. Monte Carlo studies of polymer
  network formation. \emph{The Journal of Chemical Physics} \textbf{2002},
  \emph{117}, 6300--6307\relax
\mciteBstWouldAddEndPuncttrue
\mciteSetBstMidEndSepPunct{\mcitedefaultmidpunct}
{\mcitedefaultendpunct}{\mcitedefaultseppunct}\relax
\EndOfBibitem
\bibitem[Leung and Eichinger(1984)Leung, and Eichinger]{Leung}
Leung,~Y.~K.; Eichinger,~B.~E. Computer simulation of end-linked elastomers. I.
  Trifunctional networks cured in the bulk. \emph{J. Chem. Phys.}
  \textbf{1984}, \emph{80}, 3877--3884\relax
\mciteBstWouldAddEndPuncttrue
\mciteSetBstMidEndSepPunct{\mcitedefaultmidpunct}
{\mcitedefaultendpunct}{\mcitedefaultseppunct}\relax
\EndOfBibitem
\bibitem[Grest \latin{et~al.}(1992)Grest, Kremer, and Duering]{Grest}
Grest,~G.~S.; Kremer,~K.; Duering,~E.~R. Kinetics of end crosslinking in dense
  polymer melts. \emph{Europhys. Lett.} \textbf{1992}, \emph{19},
  195--200\relax
\mciteBstWouldAddEndPuncttrue
\mciteSetBstMidEndSepPunct{\mcitedefaultmidpunct}
{\mcitedefaultendpunct}{\mcitedefaultseppunct}\relax
\EndOfBibitem
\bibitem[Gilra \latin{et~al.}(2000)Gilra, Cohen, and Panagiotopoulos]{Gilra}
Gilra,~N.; Cohen,~C.; Panagiotopoulos,~A.~Z. A Monte Carlo study of the
  structural properties of end-linked polymer networks. \emph{J. Chem. Phys.}
  \textbf{2000}, \emph{112}, 6910--6916\relax
\mciteBstWouldAddEndPuncttrue
\mciteSetBstMidEndSepPunct{\mcitedefaultmidpunct}
{\mcitedefaultendpunct}{\mcitedefaultseppunct}\relax
\EndOfBibitem
\bibitem[Schwenke and Lang(2011)Schwenke, and Lang]{Schwenke2011}
Schwenke,~K.; Lang,~J.-U.,~M.~Sommer On the Structure of Star-Polymer Networks.
  \emph{Macromolecules} \textbf{2011}, \emph{44}, 9464--9472\relax
\mciteBstWouldAddEndPuncttrue
\mciteSetBstMidEndSepPunct{\mcitedefaultmidpunct}
{\mcitedefaultendpunct}{\mcitedefaultseppunct}\relax
\EndOfBibitem
\bibitem[Lang \latin{et~al.}(2012)Lang, Schwenke, and Sommer]{Lang2012a}
Lang,~M.; Schwenke,~K.; Sommer,~J.-U. {Short cyclic structures in polymer model
  networks: A test of mean field approximation by monte carlo simulations}.
  \emph{Macromolecules} \textbf{2012}, \emph{45}, 4886--4895\relax
\mciteBstWouldAddEndPuncttrue
\mciteSetBstMidEndSepPunct{\mcitedefaultmidpunct}
{\mcitedefaultendpunct}{\mcitedefaultseppunct}\relax
\EndOfBibitem
\bibitem[Cai \latin{et~al.}(2015)Cai, Panyukov, and Rubinstein]{Cai2015}
Cai,~L.~H.; Panyukov,~S.; Rubinstein,~M. Hopping diffusion of nanoparticles in
  polymer matrices. \emph{Macromolecules} \textbf{2015}, \emph{48},
  847--862\relax
\mciteBstWouldAddEndPuncttrue
\mciteSetBstMidEndSepPunct{\mcitedefaultmidpunct}
{\mcitedefaultendpunct}{\mcitedefaultseppunct}\relax
\EndOfBibitem
\bibitem[Lang \latin{et~al.}(2001)Lang, Michalke, and Kreitmeier]{Lang2001}
Lang,~M.; Michalke,~W.; Kreitmeier,~S. A statistical model for the length
  distribution of meshes in a polymer network. \emph{The Journal of Chemical
  Physics} \textbf{2001}, \emph{114}, 7627--7632\relax
\mciteBstWouldAddEndPuncttrue
\mciteSetBstMidEndSepPunct{\mcitedefaultmidpunct}
{\mcitedefaultendpunct}{\mcitedefaultseppunct}\relax
\EndOfBibitem
\bibitem[Kilb(1958)]{Kilb1958}
Kilb,~R.~W. Dilute Gelling Systems. I. The Effect of Ring Formation on
  Gelation. \emph{The Journal of Chemical Physics} \textbf{1958}, \emph{62},
  969--971\relax
\mciteBstWouldAddEndPuncttrue
\mciteSetBstMidEndSepPunct{\mcitedefaultmidpunct}
{\mcitedefaultendpunct}{\mcitedefaultseppunct}\relax
\EndOfBibitem
\bibitem[Suematsu(1998)]{Suematsu1998}
Suematsu,~K. Theory of gel point in real polymer solutions. \emph{European
  Physical Journal B} \textbf{1998}, \emph{6}, 93--100\relax
\mciteBstWouldAddEndPuncttrue
\mciteSetBstMidEndSepPunct{\mcitedefaultmidpunct}
{\mcitedefaultendpunct}{\mcitedefaultseppunct}\relax
\EndOfBibitem
\bibitem[Lang \latin{et~al.}(2005)Lang, G{\"{o}}ritz, and
  Kreitmeier]{Lang2005a}
Lang,~M.; G{\"{o}}ritz,~D.; Kreitmeier,~S. {Intramolecular reactions in
  randomly end-linked polymer networks and linear (Co)polymerizations}.
  \emph{Macromolecules} \textbf{2005}, \emph{38}, 2515--2523\relax
\mciteBstWouldAddEndPuncttrue
\mciteSetBstMidEndSepPunct{\mcitedefaultmidpunct}
{\mcitedefaultendpunct}{\mcitedefaultseppunct}\relax
\EndOfBibitem
\bibitem[Wang \latin{et~al.}(2017)Wang, Lin, Johnson, and Olsen]{Wang2017}
Wang,~R.; Lin,~T.-S.; Johnson,~J.~A.; Olsen,~B.~D. Kinetic Monte Carlo
  Simulation for Quantification of the Gel Point of Polymer Networks. \emph{ACS
  Macro Letters} \textbf{2017}, \emph{4}, 1414--1419\relax
\mciteBstWouldAddEndPuncttrue
\mciteSetBstMidEndSepPunct{\mcitedefaultmidpunct}
{\mcitedefaultendpunct}{\mcitedefaultseppunct}\relax
\EndOfBibitem
\bibitem[Lang \latin{et~al.}(2003)Lang, G{\"o}ritz, and Kreitmeier]{Lang2003}
Lang,~M.; G{\"o}ritz,~D.; Kreitmeier,~S. {Length of subchains and chain ends in
  cross-linked polymer networks}. \emph{Macromolecules} \textbf{2003},
  \emph{36}, 4646--4658\relax
\mciteBstWouldAddEndPuncttrue
\mciteSetBstMidEndSepPunct{\mcitedefaultmidpunct}
{\mcitedefaultendpunct}{\mcitedefaultseppunct}\relax
\EndOfBibitem
\bibitem[Suematsu(2002)]{Suematsu2002}
Suematsu,~K. Recent Progress in Gel Theory: Ring, Excluded Volume, and
  Dimension. \emph{Advances in Polymer Science} \textbf{2002}, \emph{156},
  137--214\relax
\mciteBstWouldAddEndPuncttrue
\mciteSetBstMidEndSepPunct{\mcitedefaultmidpunct}
{\mcitedefaultendpunct}{\mcitedefaultseppunct}\relax
\EndOfBibitem
\bibitem[Lang \latin{et~al.}(2001)Lang, Michalke, and S.]{Lang2001b}
Lang,~M.; Michalke,~W.; S.,~K. Optimized Decomposition of Simulated Polymer
  Networks Into Meshes. \emph{Macromolecular Theory and Simulations}
  \textbf{2001}, \emph{10}, 204--208\relax
\mciteBstWouldAddEndPuncttrue
\mciteSetBstMidEndSepPunct{\mcitedefaultmidpunct}
{\mcitedefaultendpunct}{\mcitedefaultseppunct}\relax
\EndOfBibitem
\bibitem[Kricheldorf \latin{et~al.}(2001)Kricheldorf, B{\"o}hme, and
  Schwarz]{Kricheldorf2001}
Kricheldorf,~H.~R.; B{\"o}hme,~S.; Schwarz,~G. Macrocycles. 17. The role of
  cyclization in kinetically controlled polycondensations. 2. Polyamides.
  \emph{Macromolecules} \textbf{2001}, \emph{34}, 8879--8885\relax
\mciteBstWouldAddEndPuncttrue
\mciteSetBstMidEndSepPunct{\mcitedefaultmidpunct}
{\mcitedefaultendpunct}{\mcitedefaultseppunct}\relax
\EndOfBibitem
\bibitem[Miller and Macosko(1976)Miller, and Macosko]{Miller1976}
Miller,~D.~R.; Macosko,~C.~W. {A New Derivation of Post Gel Properties of
  Network Polymers}. \emph{Macromolecules} \textbf{1976}, \emph{9},
  206--211\relax
\mciteBstWouldAddEndPuncttrue
\mciteSetBstMidEndSepPunct{\mcitedefaultmidpunct}
{\mcitedefaultendpunct}{\mcitedefaultseppunct}\relax
\EndOfBibitem
\bibitem[Lang(2004)]{Lang2004}
Lang,~M. {Bildung und Struktur von polymeren Netzwerken, Dissertation}. Ph.D.\
  thesis, Universit{\"{a}}t Regensburg, 2004\relax
\mciteBstWouldAddEndPuncttrue
\mciteSetBstMidEndSepPunct{\mcitedefaultmidpunct}
{\mcitedefaultendpunct}{\mcitedefaultseppunct}\relax
\EndOfBibitem
\bibitem[Wang \latin{et~al.}(2016)Wang, Alexander-Katz, Johnson, and
  Olsen]{Wang2016}
Wang,~R.; Alexander-Katz,~A.; Johnson,~J.~A.; Olsen,~B.~D. Universal Cyclic
  Topology in Polymer Networks. \emph{Physical Review Letters} \textbf{2016},
  \emph{116}, 188302\relax
\mciteBstWouldAddEndPuncttrue
\mciteSetBstMidEndSepPunct{\mcitedefaultmidpunct}
{\mcitedefaultendpunct}{\mcitedefaultseppunct}\relax
\EndOfBibitem
\bibitem[Adam and Delsanti(1983)Adam, and Delsanti]{Adam1983}
Adam,~M.; Delsanti,~M. {Viscosity and longest relaxation time of semi-dilute
  polymer solutions. I. Good solvent}. \emph{Journal de Physique}
  \textbf{1983}, \emph{44}, 1185--1193\relax
\mciteBstWouldAddEndPuncttrue
\mciteSetBstMidEndSepPunct{\mcitedefaultmidpunct}
{\mcitedefaultendpunct}{\mcitedefaultseppunct}\relax
\EndOfBibitem
\bibitem[Lang and Sommer(2010)Lang, and Sommer]{Lang2010}
Lang,~M.; Sommer,~J.-U. Analysis of Entanglement Length and Segmental Order
  Parameter in Polymer Networks. \emph{Physical Review Letters} \textbf{2010},
  \emph{104}, 177801\relax
\mciteBstWouldAddEndPuncttrue
\mciteSetBstMidEndSepPunct{\mcitedefaultmidpunct}
{\mcitedefaultendpunct}{\mcitedefaultseppunct}\relax
\EndOfBibitem
\bibitem[Lang(2013)]{Lang2013}
Lang,~M. {Monomer fluctuations and the distribution of residual bond
  orientations in polymer networks}. \emph{Macromolecules} \textbf{2013},
  \emph{46}, 9782--9797\relax
\mciteBstWouldAddEndPuncttrue
\mciteSetBstMidEndSepPunct{\mcitedefaultmidpunct}
{\mcitedefaultendpunct}{\mcitedefaultseppunct}\relax
\EndOfBibitem
\bibitem[Cohen-Addad(1973)]{CohenAddad}
Cohen-Addad,~J.~P. Effect of the anisotropic chain motion in molten polymers:
  the solidlike contribution of the nonzero average dipolar coupling to NMR
  signals theoretical description. \emph{J. Chem. Phys.} \textbf{1973},
  \emph{60}, 2440--2453\relax
\mciteBstWouldAddEndPuncttrue
\mciteSetBstMidEndSepPunct{\mcitedefaultmidpunct}
{\mcitedefaultendpunct}{\mcitedefaultseppunct}\relax
\EndOfBibitem
\bibitem[Saalw{\"{a}}chter(2007)]{Saalwaechter}
Saalw{\"{a}}chter,~K. Proton multiple-quantum NMR for the study of chain
  dynamics and structural constraints in polymeric soft materials. \emph{Prog.
  Nucl. Magn. Res. Spectr.} \textbf{2007}, \emph{51}, 1--35\relax
\mciteBstWouldAddEndPuncttrue
\mciteSetBstMidEndSepPunct{\mcitedefaultmidpunct}
{\mcitedefaultendpunct}{\mcitedefaultseppunct}\relax
\EndOfBibitem
\end{mcitethebibliography}
\end{document}